\documentclass[12pt]{article}
\usepackage{graphicx}
\usepackage{hyperref}
\usepackage[all,knot,arc,import,poly,cmtip]{xy}
\usepackage{amsmath}
\usepackage{amscd,amssymb}
\usepackage{xcolor}

\newcommand{\cA}{\mathcal{A}}
\newcommand{\cE}{\mathcal{E}}

\newcommand{\cH}{\mathcal{H}}

\newcommand{\cG}{\mathcal{G}}

\newcommand{\xX}{\textsf{X}}
\newcommand{\yY}{\textsf{Y}}
\newcommand{\wW}{\textsf{W}}
\newcommand{\zZ}{\textsf{Z}}
\newcommand{\MM}{\textsf{M}}
\newcommand{\nN}{\textsf{N}}

\def\cR{{\mathcal R}}

\def\cA{{\mathcal A}}

\def\cC{{\mathcal C}}
\def\cK{{\mathcal K}}

\def\cH{{\mathcal H}}

\def\cL{{\mathcal L}}
\def\cE{{\mathcal E}}

\def\cY{{\mathcal Y}}
\def\cG{{\mathcal G}}

\def\mD{{\mathfrak D}}

\newtheorem{example}{Example}[section]

\newcommand{\beq}{\begin{eqnarray}}
\newcommand{\eeq}{\end{eqnarray}}
\numberwithin{equation}{section}

\begin{document}


\begin{center}
{\large\bf Chern-Simons Invariants on Hyperbolic Manifolds and Topological 
Quantum Field Theories}
\end{center}

\vspace{0.1in}

\begin{center}
{\large
L. Bonora $^{(a)}$
\footnote{E-mail: bonora@sissa.it},
A. A. Bytsenko $^{(b)}$
\footnote{E-mail: aabyts@gmail.com},
A. E. Gon\c{c}alves $^{(b)}$
\footnote{E-mail: aedsongoncalves@gmail.com}}

\vspace{5mm}
$^{(a)}$
{\it International School for Advanced Studies (SISSA/ISAS) \\
Via Bonomea 265, 34136 Trieste and INFN, Sezione di Trieste, Italy}

\vspace{0.2cm}
$^{(b)}$
{\it
Departamento de F\'{\i}sica, Universidade Estadual de
Londrina\\ Caixa Postal 6001,
Londrina-Paran\'a, Brazil}



\end{center}

\vspace{0.1in}

\begin{abstract}
We derive formulas for the classical Chern-Simons invariant of irreducible 
$SU(n)$-flat connections on negatively curved locally symmetric three-manifolds.
We determine the condition for which the theory remains consistent (with  basic physical 
principles). We show that a connection between holomorphic values of Selberg-type functions 
at point zero, associated with R-torsion of the flat bundle, and twisted Dirac operators 
acting on negatively curved manifolds, can be interpreted by means of the Chern-Simons 
invariant. On the basis of Labastida-Mari\~{n}o-Ooguri-Vafa conjecture
we analyze a representation of the Chern-Simons quantum partition function (as a generating
series of quantum group invariants) in the form of an infinite product weighted by 
S-functions and Selberg-type functions. We consider the case of links and a knot and
use the Rogers approach to discover certain symmetry and modular form identities.
\end{abstract}

\vspace{0.1in}

\begin{flushleft}
PACS \, 11.15. Yc (Chern-Simons gauge theory), 02.10. Kn (Knot theory), \\
02.20. Uw (Quantum groups)

\vspace{0.3in}
October 2016
\end{flushleft}

\newpage

\tableofcontents


\section{Introduction}
\label{Introduction}

The Chern-Simons theory is one of the two archetypal field theories in physics, 
together with Yang-Mills theory, that describe the interaction of gauge fields. In 3d, the 
dimension we wish to consider in this paper, the dynamics is so constrained as to leave room 
only for non-dynamical (topological) correlators.  The corresponding topological quantum 
field theory was defined and developed by Witten \cite{Witten88} and Reshetikhin-Turaev 
\cite{RT}, and applied to the mathematical theory of knots and links in three-dimensional 
manifolds. From 
a field theory point of view, the observables of the Chern-Simons theory are correlators of Wilson 
lines (beside the partition function). In this paper we will analyze some properties of the latter. 
 
A well-known characteristic of the Chern-Simons path integral is
that its well-definiteness is connected to the properties of a topological invariant in 
four-dimensional manifolds. Such invariant is related to the Chern-Simons form via the 
transgression formula. The invariant of a four-manifold 
in the topological field theory involves its signature and Euler characteristic (see for example 
\cite{CY}). The Chern character allows one to map the analytical Dirac index in terms of 
K-theory classes into a topological index, which can be expressed in terms of cohomological 
characteristic classes.
This results in a connection between the Chern-Simons action and the Atiyah-Singer index theorem.
Such a connection will be used in this paper in order to determine the Chern-Simons invariant 
of irreducible 
$SU(n)$-flat connections on negatively curved locally symmetric three-manifolds. Indeed a 
critical point of the Chern-Simons functional is just a flat connection; it corresponds to 
a representation of the fundamental group $\pi_1(\xX)$ associated to a three-manifold $\xX$. 
The value of the Chern-Simons functional at a critical point can be regarded as a topological
invariant of a pair $(\xX, \rho)$, where $\rho$ is a representation of $\pi_1(\xX)$. Due to
a well known adiabatic argument, knowing these invariants allows to compute the  
partition function. 

On the other hand, the Chern-Simons partition function is a generating 
series of quantum group invariants weighted by S-functions. Recall that the Chern-Simons 
theory has been conjectured to be equivalent to a topological string theory $1/N$ expansion 
in physics. The Chern-Simons/topological string duality conjecture identifies
the generating function of Gromov-Witten invariants as Chern-Simons knot invariants \cite{OV}.  
The existence of a sequence of integer invariants is conjectured \cite{OV, LMV} in a similar 
spirit to the Gopakumar-Vafa setting \cite{GV}. This provides essential evidence of the
duality between Chern-Simons theory and topological string theory. Such an integrality conjecture
is called the LMOV conjecture. In the context of this conjecture we derive a new 
representation of the Chern-Simons quantum partition function in the form of an infinite product
in terms of Selberg-type functions.

Deeply related with the content of this paper is the problem of anomalies. 
In field theory anomalies may prevent the path integral from being well defined. In the case 
of CS in three-dimensional manifolds there are no local anomalies, but there may be global 
anomalies. To guarantee their absence one must restrict to integer
values the (suitably normalized) coupling appearing in front of the action. For this reason it is 
of utmost importance to know the value of the Chern-Simons invariant in any given space-time. 
Strictly connected with this is the issue of existence of fermionic path integrals 
(fermion determinants) in three-dimensional manifolds.  There are
also other indeterminacies in this theory when links and knots are involved, related to the 
evaluation of overlapping Wilson loops. The problem of such framing anomalies 
was pointed out and solved by 
Witten \cite{Witten89}.

{\bf Our key results.}
More specifically the content and main results of our paper are as follows.

-- In Sect. \ref{Abelian} we derive the formula for the Chern-Simons invariant of  
irreducible $SU(n)$-flat connections on a locally symmetric manifold of non-positive 
sectional curvature. For this Chern-Simons invariant our result, Eq. (\ref{sequential}), 
determines the condition for which the quantum field theory is consistent. 
The results of Sect. \ref{Abelian} are preparatory for the generalization of the Chern-Simons 
invariant to the case of other manifolds ($X = S^3/\Gamma$, for example) and of 
non-trivial $U(n)$-bundle over $X$ (Sect. \ref{Unitary}).

-- In Sect. \ref{Unitary} $\xX = \Gamma\backslash \overline{\xX}$ with $\overline{\xX}$ is
a globally symmetric space of non-compact type and $\Gamma$ a discrete, torsion-free, 
co-compact subgroup of orientation-preserving isometries. $\xX$ inherits a locally symmetric
Riemannian metric $g$ of non-positive sectional curvature. 
If ${\mD}: C^{\infty}(\textsf{X}, V)\rightarrow C^{\infty}(\textsf{X},V)$ is a differential
operator acting on the sections of the vector bundle $V$, then
${\mathfrak D}$ can be extended canonically to a differential operator
${\mathfrak D}_{\varphi}: C^{\infty}(\textsf{X},V\otimes F)\rightarrow
C^{\infty}(\textsf{X},V\otimes{F})$, uniquely
characterized by the property that ${\mathfrak D}_{\varphi}$ is
locally isomorphic to ${\mathfrak D}\otimes \cdots \otimes {\mathfrak
D}$\,\,\, (${\rm dim}\,{F}$ times) \cite{Moscovici89}.
We show that a connection between holomorphic values of Selberg-type functions at point zero, 
associated with R-torsion of the flat bundle, and twisted Dirac operators $\mD_\varphi$
on negatively curved locally symmetric spaces, can be interpreted by means of the 
Chern-Simons invariant. This leads to our main result,  {Eq. (\ref{mainfinal})}. 
We  also briefly describe the possibility  
to derive the Chern-Simons invariant for locally symmetric spaces of higher rank 
in terms of the spectral function ${\cR}(s; \varphi)$.

-- The quantum $\mathfrak{sl}_N$ invariant in the case of links and a knot 
is analyzed in Sect. \ref{QuantumCS}. On the base of LMOV conjecture we derive a new 
representation of the Chern-Simons quantum partition function in the form of an infinite product
in terms of Selberg-type functions. In addition, we discuss the symmetry and modular 
form properties of infinite product formulas.

\section{Chern-Simons invariants for negatively curved manifolds}
\label{ClassicCS}

\subsection{Flat connections and gauge bundles}
\label{Abelian}

{\bf Flat connections on fibered hyperbolic manifolds.}
The Chern character allows one to map the analytical Dirac index
in terms of K-theory classes into a topological index which can be
expressed in terms of cohomological characteristic classes. This
results in a connection between the Chern-Simons action
and the celebrated Atiyah-Singer index theorem.
The goal of this section is to  use this fact in order to present explicit
formulas for the Chern classes and gauge Chern-Simons invariant of an irreducible $SU(n)$-flat 
connection on real compact hyperbolic three-manifolds.

Let ${P}=\textsf{X}\times{G}$ be a trivial principal bundle over $\textsf{X}$ 
with the gauge group ${G}=SU(n)$ and let $\Omega^1(\textsf{X};{\mathfrak g})$ be the space
of all connections on ${P}$; this space is an affine space of one-forms on $\textsf{X}$ with 
values in the Lie algebra ${\mathfrak g}$ of ${G}$. Let  
$\cA_{\textsf{\xX}}= \Omega^1(\textsf{X};{\mathfrak g})$ be the space of connections and 
$\cA_{\textsf{\xX}, F}=\{A\in \cA_{\textsf{X}}|F_A= dA+A\wedge A=0\}$ be the space 
of flat connections on $P$.
Then the gauge transformation group $\cG_{\xX}\cong C^\infty(\xX, G)$ acts on 
$\cA_{\textsf{X}}$ via pull-back: ${\mathfrak g}^\ast A = {\mathfrak g}^{-1}A{\mathfrak g} + 
{\mathfrak g}^{-1}d{\mathfrak g}$, ${\mathfrak g}\in \cG_{\xX}$, 
$A\in \cA_{\textsf{X}}$. This action preserves $\cA_{\textsf{\xX},F}$.

It is known that the Chern-Simons invariant is a real valued function on the space of
connections $\cA_{\textsf{X}}$ on a trivial principal bundle over a oriented 
three-manifold, and  it is given by
\begin{equation}
CS(A)=
\frac 1{8\pi^2}\int_{\textsf{X}}\mbox{Tr} \left(A\wedge dA+ \frac 23\,A
\wedge A\wedge A\right)\,.
\label{CS0}
\end{equation}
A well-known formula related to the $CS$ integrand is:
\begin{equation}
d \,{\rm Tr}\, (A\wedge dA+ (2/3)A\wedge A\wedge A) = {\rm
Tr}\left(F_{A}\wedge F_{A}\right)\,.
\label{AA}
\end{equation}
Eq. (\ref{AA}) provides another description of the Chern-Simons invariant. Indeed, let
$\xX$ be a closed manifold, and $\MM$ be an oriented four-manifold with $\partial\MM = \xX$.
We denote an extension of $A$ over $\textsf{M}$ by ${\underline A}$.
Then the Stokes' theorem gives
\begin{equation}
{CS}({\underline A})=
\frac{1}{8\pi^2}\int_{\textsf{M}}\mbox{Tr}\left(F_{{\underline A}}
\wedge F_{{\underline  A}}\right).
\label{CS0'}
\end{equation}
 
We note the difference of the Chern-Simons invariants under the 
action of the gauge transformation group $\cG_{\xX}$. In the case 
$A \in \cA_\xX$ and ${\mathfrak g} \in \cG_{\xX}$ the following formula holds \cite{Nishi}
\begin{equation}
CS({\mathfrak g}^*A) - CS(A) = \frac{1}{8\pi^2}\int_{\partial \xX}{\rm Tr}
(A\wedge d{\mathfrak g}{\mathfrak g}^{-1}) - \frac{1}{24\pi^2}
\int_{\xX}{\rm Tr} ({\mathfrak g}^{-1}d{\mathfrak g}\wedge{\mathfrak g}^{-1}
d{\mathfrak g}\wedge {\mathfrak g}^{-1}d{\mathfrak g}).
\label{WZW}
\end{equation}
The last term in Eq. (\ref{WZW}) is known as the Wess-Zumino-Witten term; the 
integrand of this term represents the generator of 
$H^3(G; {\mathbb Z})\cong {\mathbb Z}$.
If the manifold $\xX$ is closed, i.e. $\partial \xX = \emptyset$, 
the Wess-Zumino-Witten term takes its value in $\mathbb Z$. In this case the function
$CS: \cA_\xX/\cG_\xX \rightarrow {\mathbb R}/{\mathbb Z}$ is well defined. On the other hand, 
when $\partial \xX \neq \emptyset$, although the Chern-Simons invariant does not give a 
well-defined function on the space $\cA_\xX/\cG_\xX$ (with values in 
${\mathbb R}/{\mathbb Z}$), one can regard it as the section of a certain line bundle
over the moduli space of connections $\cA_\xX/\cG_\xX$.

Let us consider now the moduli space of flat connections ${\mathcal X}_\xX =
\cA_{\xX, F}/\cG_\xX$ on $\xX$. This space has an alternative topological 
description: the holonomies of the parallel transport of flat connections on $P$ give 
the identification of ${\mathcal X}_\xX$ with the space of conjugacy classes of 
representations of  $\pi_1(\xX)$ into $G$, since any principal $G$-bundle $P$ over 
a compact oriented three-manifold $\xX$ is trivial \cite{Nishi}. 
We use the notation $CS(\rho) := CS(A_\rho)$
for a representation $\rho$ of $\pi_1(\xX)$, where $A_\rho$ is a flat connection 
corresponding to a representation $\rho$. This gives a topological invariant
for a pair $(\xX, \rho)$. Since the Chern-Simons invariant is additive with respect to 
the sum of representations, we have:
\begin{equation}
CS(\rho_1\oplus \rho_2) = CS(\rho_1) + CS(\rho_2)\,. 
\end{equation}

Let $\xX$ be a compact oriented hyperbolic three-manifold, and $\rho$ be an irreducible 
representation of $\pi_1(\xX)$ into $SU(n)$. Denote the corresponding flat vector 
bundle by $E_\rho$, and a flat extension of $A_\rho$ over $\MM$ ($\partial\MM = \xX$)
corresponding to $\rho$ by $\underline{A}_\rho$.
The second Chern
character ${\rm ch}_2({{\underline E}}_{\rho})\, = CS({\underline A})$
of ${ {\underline  E}}_{\rho}$ can be
expressed in terms of the first and second Chern classes
\begin{equation}
{\rm ch}_2({ {\underline  E}}_{\rho})= (1/2)c_1({
{\underline E}}_{\rho})^2 - c_2({ {\underline   E}}_{\rho})\,,
\label{ch2}
\end{equation}
while the Chern character is given by
\begin{equation}
{\rm ch}({ {\underline   E}}_{\rho})  =  {\rm rank}\,
{ {\underline   E}}_{\rho}+
c_1({ {\underline   E}}_{\rho})
+{\rm ch}_2({ {\underline   E}}_{\rho})
={\rm dim}\,\rho +c_1({ {\underline   E}}_{\rho})+
{\rm ch}_2({ {\underline   E}}_{\rho})\,.
\label{chE}
\end{equation}

The crucial point in our calculation is the
Atiyah-Patodi-Singer result for manifold with boundary \cite{Atiyah77,Atiyah78,Atiyah79}:
the Dirac index is given by
\begin{equation}
{\bf Index}\, {\mathfrak D}_{\rho} =\int_\textsf{M}
{\rm ch}
({ {\underline   E}}_{\rho})
{\widehat  A}(\textsf{M})-\frac{1}{2}(\eta(0,{\mathfrak D}_{\rho})+
h(0,{\mathfrak D}_{\rho}))\,,
\label{Index}
\end{equation}
Here ${\widehat A}(\textsf{M}) \equiv {\widehat A}(\Omega(\textsf{M}))$-genus is the usual
polynomial in terms of the Riemannian curvature $\Omega(\textsf{M})$
of a four-manifold $\textsf{M}$ with boundary
$\partial \textsf{M} = \textsf{X}$. It
is given by ${\widehat  A}(\textsf{M})= \left({\rm det}
\left(\frac{\Omega(\textsf{M})/4\pi}{{\rm sinh}
\, \Omega(\textsf{M})/4\pi} \right)\right)^{1/2}$ 

$= 1- (1/24)p_1({\textsf{M}})$, where
$p_1(\MM) \equiv p_1(\Omega({\textsf{M}}))$ is the first Pontrjagin class.
$h(0,{\mathfrak D}_{\rho})$ is the dimension of the space of harmonic
spinors on $\textsf{X}$ ($h(0,{\mathfrak D}_{\rho})
={\rm dim}{\rm Ker}\,{\mathfrak D}_{\rho}$ =
multiplicity of the 0-eigenvalue of ${\mathfrak  D}_{\rho}$ acting
on $\textsf{X}$ with coefficients in $\rho$).
The $\eta$-invariant was introduced by Atiyah, Patodi, and Singer 
\cite{Atiyah77,Atiyah78,Atiyah79} treating index theory on even dimensional manifolds with 
boundary and it first appears there as a boundary correction in the usual local index formula. 
Let as before $\xX$ be a closed odd dimensional spin manifold (which in their index theorem is 
the boundary of an even dimensional spin manifold). 
$\eta_{\xX}(s, \mD):= \eta(s, \mD_{\rho=trivial})$ is 
analytic in $s$ and has a meromorphic continuation to $s\in {\mathbb C}$; it is regular 
at $s=0$, and its value there is the $\eta$-invariant.
The result (\ref{Index}) holds for any Dirac operator on a $Spin^{\mathbb C}$ manifold coupled 
to a vector bundle with connection (the metric of manifolds is supposed to be a product near 
the boundary). One can attach an $\eta$-invariant to any operator of Dirac type on a compact 
Riemannian manifold of odd dimension. (On even dimension manifolds Dirac operators 
have symmetric spectrum and, therefore, trivial $\eta$-invariant.)

 {The $\eta$ is a spectral invariant which measures the symmetry of the spectrum of 
an operator $\mD$, and admits a meromorphic extension to the whole $s$-plane, with 
at most simple poles at $(dim \xX-k)/(ord\, \mD)$ (k=0,1,2,...) and locally computable 
residues. For $\xX$ a compact oriented (4n-1)-dimensional Riemannian manifold of 
constant negative curvature, a remarkable  formula relating $\eta$ to the 
closed geodesics on $\xX$ has been proved in \cite{Milson}. Citing \cite{Moscovici89}}, 
the appropriate class 
of Riemannian manifolds for which a result of this type can be expected is that 
of non-positively curved locally symmetric manifolds, while the class of self-adjoint 
operators whose eta invariants are interesting to compute is that of Dirac-type 
operators, even with additional coefficients in locally flat bundles. It is one 
of the purpose of this paper to formulate and prove for Chern-Simons invariants (2.12) below
such an extension as the one in \cite{Milson} (see Sect. 2.4).

{For a trivial representation $\rho$ one can choose a trivial 
flat connection ${\underline A}$;
then $F_{\underline A} = 0$. For this choice using Eqs. (\ref{ch2}) and 
(\ref{chE}) we have
}
\begin{eqnarray}
\int_\textsf{M} {\rm ch}({ {\underline E}}_{\rho}){\widehat  A}(\textsf{M})
& = & \int_\textsf{M} ({\rm rank}{\underline E}_\rho +
c_1({ {\underline   E}}_{\rho})+
{\rm ch}_2({ {\underline  E}}_{\rho}))
(1-\frac{1}{24}p_1(\textsf{M}))
\label{E1}
\\
& = &
- \frac{1}{8\pi^2}\int_\textsf{M} {\rm Tr} (F_{{\underline A}_\rho}
\wedge F_{{\underline A}_\rho}) - \frac{{\rm dim}\,\rho}{24}\int_\textsf{M}
p_1(\textsf{M})
\nonumber \\
 {\rm while} \,\,\,\,\,\, \int_\textsf{M} {\rm ch}({\underline  E})
{\widehat  A}(\textsf{M})
& = & - \frac{{1}}{24}\int_\textsf{M} p_1(\textsf{M}) {\bf\, 
-\frac{1}{2}\eta (0, {\mathfrak D})} \,.
\label{E2}
\end{eqnarray}
{Note that in this formula the zero eigenvalue multiplicity of $\mD_\rho$ acting on X with   
coefficients in $\rho$ has been excluded  
($s \in  i\,Spec^\prime({\mathfrak D}), Spec^\prime(\mD) \equiv Spec(\mD) - \{ 0 \}$). 
}{From Eqs. (\ref{E1}) and (\ref{E2}) we obtain:}
\begin{equation}
{\bf Index}\,{\mathfrak D}_{\rho}  - {\rm dim}\,\rho \cdot {\bf Index}\,{\mathfrak D}
=  CS({{\underline A}_{\rho}}) - \frac{1}{2}(\eta (0, {\mathfrak D}_\rho) -
{\rm dim} \rho \cdot \eta (0, {\mathfrak D})).
\label{Index2}
\end{equation}
Then the Chern-Simons invariant can be derived from Eq. (\ref{Index2}),
\begin{equation}
CS({\underline A}_{\rho})  = (1/2)(\eta (0, {\mathfrak D})
- {\rm dim} \rho \cdot\eta (0, {\mathfrak D}_\rho))\,\, + \,\,
{\rm modulo} \,\,\,\, \mathbb Z \,.
\label{CS1}
\end{equation}

\subsection{The Chern-Simons-- and the $\eta$--invariants}

{\bf Gluing properties for $\eta$.}
 Let  $\MM^{\,\prime}$ and $\MM^{\,\prime\prime}$ be oriented manifolds and let  
$-\MM^{\,\prime}$ be the 
manifold with opposite orientation with respect to  $\MM^{\,\prime\prime}$. If  
$\MM^{\,\prime}$ and $\MM^{\,\prime\prime}$ have a common boundary we can glue them along it and
form a new oriented and closed manifold $\nN= \MM^\prime\bigcup (-\MM^{\,\prime\prime})$. 
Then the following gluing formula holds:
\begin{equation}
\exp(i\pi\eta(\MM^{\,\prime})) = \exp(i\pi\eta(\MM^{\,\prime\prime}))\cdot
\exp(i\pi\eta(\nN)),
\label{gluing}
\end{equation}
This property is instrumental for the main conclusion of this part of the paper contained in
{r\'{e}sum\'{e}:}
\label{Resume}
A critical point of the Chern-Simons functional
is just a flat connection, and it corresponds to a representation of the fundamental
group $\pi_1(\xX)$. Thus, the value of this functional at a critical point can be regarded 
as a topological invariant of a pair $(\xX, \rho)$, where $\rho$ is a representation of
$\pi_1(\xX)$. This is the Chern-Simons invariant of a flat connection on $\xX$.
Taking into account a pair $(\MM, \,\partial\MM = \xX)$
we have derived Eq. {\rm (\ref{CS1})} for the Chern-Simons invariants of irreducible 
$SU(n)$-flat connections on a locally symmetric manifolds of non-positive section curvature.

By making use Eq. {\rm (\ref{CS1})} one can rewrite the multiplicative structure of 
eta invariants, associated with twisted Dirac operators $\mD_\rho$
\footnote{
Cf. Eq. (\ref{gluing}): the $\eta$-invariant of locally symmetric manifolds of non-positive
curvature can be expressed as spectral values of zeta-functions constructed out of the 
periodic geodesics \cite{Moscovici89,Moscovici91}. In this case a necessary regularization 
can be given by the geodesic spectrum.
For the time being we assumed the other regularization for the $\eta$-invariant, which involves 
the ``dual'' data, namely the spectrum of the Laplace operator associated to the 
metric \cite{Moscovici91}. 
}{
\begin{equation}
\exp \left(i2\pi  CS({\underline A}_{\rho})\right) =
\exp \left(i\pi \eta (0, {\mathfrak D}_\rho)\right) \cdot
\exp \left(-i\pi \, {\rm dim} \rho \cdot \eta (0, {\mathfrak D})\right)\,.
\end{equation}}

{\bf Dirac operators on locally symmetric spaces of rank one.}
One can repeat the technique and arguments discussed in Sect. \ref{Abelian} for the 
construction of the eta functions and the Chern-Simons invariant.
The Chern-Simons invariant admits a representation in terms of Selberg-type spectral function
$Z(s, \mD_\rho)$, which is a meromorphic function on $\mathbb C$ and given for
${\rm Re} (s^2)\gg 0$ \cite{Moscovici89,Moscovici91}.
${\rm log}Z(s,{\mathfrak D}_\rho)$ has a meromorphic continuation given by the identity
\begin{equation}
{\rm log}Z(s,{\mathfrak D_\rho})={\rm log}\,{\rm det}^{\prime}
\left(\frac{{\mathfrak D}_\rho- is}{{\mathfrak D}_\rho + is}\right)
+ i\pi\eta(s,{\mathfrak D}_\rho)
\mbox{,}
\label{Z1}
\end{equation}
$Z(s,{\mathfrak D}_\rho)$
satisfies the functional equation
$
Z(s,{\mathfrak D}_\rho)Z(-s,{\mathfrak D}_\rho)=
\exp \left(2\pi i \eta (s,{\mathfrak D}_\rho)\right),
$
where the ``twisted'' zeta function $Z(s, \mD_\rho)$ is meromorphic on $\mathbb C$. 
Zeta functions are given by the formulas \cite{Moscovici89}:
\begin{eqnarray}
{\rm log}\,Z(s, \mD) & = & (-1)^q \sum_{[\gamma]\in {\cE_1(\Gamma)}}
\frac{L(\gamma, \mD)}
{\vert{\rm det}(I-P_h(\gamma))\vert^{1/2}}\frac{e^{-s\ell_\gamma}}{m_\gamma}\,, 
\\
{\rm log}\,Z(s, \mD_\rho) & = & (-1)^q \sum_{[\gamma]\in {\cE_1(\Gamma)}}
{\rm Tr}\,\rho (\gamma)\frac{L(\gamma, \mD)}
{\vert{\rm det}(I-P_h(\gamma))\vert^{1/2}}\frac{e^{-s\ell_\gamma}}{m_\gamma}\,, 
\end{eqnarray}
where $\ell(\gamma)$ is the length of the closed geodesic $c_\gamma$ in the free homotopy 
class corresponding to $[\gamma]$, $m(\gamma)$ is the multiplicity of $c_\gamma$, 
$L(\gamma, \mD)$ are the Lefschitz numbers, and $P_h(\gamma)$ is the hyperbolic part of 
the linear Poincar\'{e} map $P(\gamma)$ (see for detail \cite{Moscovici89}).

Taking into account that the Dirac operator is Hermitian, and the function 
$CS ({\underline A}_{\rho})$ is real, it is possible to formulate the following result :
\begin{equation}
\exp \left(i2\pi  CS ({\underline A}_{\rho})\right) =  
\frac{Z(0, {\mathfrak D})^{\dim\, \rho}}
{Z(0, {\mathfrak D_\rho})}
=
(-1)^{{\rm dim}\,{\rm ker}({\mathfrak D}_\rho - {\rm dim\,\rho}\cdot {\mathfrak D})}
\,,
\label{mainfinal}
\end{equation}
as it follows from Eq. {\rm (\ref{CS1})}.
In particular 
$
Z(0, {\mathfrak D})^{\dim\, \rho}\cdot Z(0, {\mathfrak D_\rho})^{-1} = \pm 1
$
(\cite{Moscovici89}, Corollary 7.5). There is an indeterminacy of sign unless 
\begin{equation}
{\rm dim}\,{\rm ker}({\mathfrak D}_\rho - {\rm dim\,\rho}\cdot {\mathfrak D})
= 2n, \,\,\,\,\, n\in {\mathbb Z}\,.
\label{sequential}
\end{equation}

\subsection{Determinant line bundles}

In the previous subsections we have studied the connection of the CS invariant with the 
$\eta$-invariant. The latter, on the other hand, features also in the context of anomaly 
formulas, which are related to the geometry of the determinant line bundles. In this subsection 
we would like to recall such a connection. We have in mind in particular a three-dimensional
manifold $\xX$, as above, but it is possible to stick to a more general treatment.

In the sequel we suppose that $\yY$ is a compact $Spin^{\mathbb C}$-manifold with nonempty
boundary. The Dirac operator ${\mD}$ on a closed $Spin^{\mathbb C}$-manifold $\yY$ 
(coupled to a vector bundle with connection) is self-adjoint and has a discrete spectrum 
${\rm Spec}\,({\mD})$. We suppose that the metric of $\yY$ near the boundary has explicit 
product structure, and in a neighborhood of the boundary there is a given isometry with
$(-1, 0]\times\partial\yY$.
$\eta(s, \mD)$ is analytic for ${\rm Re}\,(s)>-2$ \cite{bismut1,bismut2}, and we set
\begin{equation}
\tau_{\yY}= \exp (i2\pi \xi_{\yY}) \equiv \exp\left(i\pi (\eta(0,{\mD})+{\rm dim}\,{\rm ker}\,
{\mD})
\right)\in{\mathbb C}\,.
\label{tau}
\end{equation}
Under a smooth variation of parameters (for example, the metric on $\yY$) the $\eta$-invariant 
jumps by integers (the general theory shows that $|\tau_{\yY}|=1$), whereas $\xi$ (mod 1) 
is smooth. 
Therefore the invariant (\ref{tau}) is defined and we have 
$\tau_{\yY} \in {\rm det}_{\partial {\yY}}^{-1}$,
where ${\rm det}_{\partial {\yY}}$ is the determinant line of the Dirac operator 
${\mD}_{\partial {\yY}}$ on the boundary.

{\bf Fiber bundles.}
Let us discuss some aspects of Dirac operators in the case of fiber bundles. 
Let $\pi: {\wW}\rightarrow \zZ$  be a smooth fiber bundle with a Riemannian metric
on the tangent bundle $T(\wW/\zZ)$, which is endowed with spin structure. Here and in the following
$\wW/\zZ$ denotes the fiber of $\wW\rightarrow \zZ$. A spin structure 
on a manifold means a spin structure on its tangent bundle, in this case the tangent bundle 
$T(\wW/\zZ)$  along the fibers. Every point in $\zZ$ determines a Dirac operator acting 
on the corresponding fiber. We will eventually identify the fiber $\partial \wW/\zZ$ with a three-dimensional
manifold $\xX$ without boundary, but for the time being we keep as general as possible.

Assume that the Riemannian metric on the fibers is a product near the boundary. 
The determinant line carries the Quillen metric and a canonical connection $\nabla$ 
\cite{bismut1} and the exponentiated $\xi$-invariant is a smooth section 
$\tau_{\wW/\zZ}: \zZ\rightarrow {\rm det}_{{\partial \wW}/\zZ}^{-1}$.


For Dirac operators coupled to complex bundles in K-theory one can 
express the Chern character of the index in terms of Chern character of a complex
vector bundle $E$ by means of the formula \cite{freed99}:
\begin{equation}
{\rm ch}\,\pi_{!}^{\wW/\zZ}([E])=\pi_{*}^{\wW/\zZ}({\widehat A}(\wW/\zZ){\rm ch}\,(E)),
\end{equation}
where $\pi_{*}$ is the pushforward map in rational cohomology.
Note that for a family of closed manifolds this is a result of Atiyah-Patodi-Singer 
(see formula (\ref{Index})).

Let the fibers $\partial\wW/\zZ$ be odd-dimensional and closed. Then the determinant line bundle 
${\rm det}\,{\mD}_{\partial\wW/\zZ}(E)$ is well-defined as a smooth line bundle
(it carries a canonical metric and connection) \cite{bismut1}. The complex
Dirac operator for the fibers $\partial\wW/\zZ$ is self-adjoint and 
there is a geometrical invariant $\tau_{\wW/\zZ}$ defined by Atiyah-Patodi-Singer,
\begin{equation}
\tau_{\wW/\zZ}(E) = \exp\left(i\pi (\eta(0, \mD)+{\rm dim}\,{\rm ker}\,
{\mD})_{\partial\wW/\zZ}\right)\,.\label{tauMN}
\end{equation}

The 2-form curvature of the determinant line bundle is \cite{freed99}
\begin{equation}
\Omega({\rm det}\,{\mD}_{\wW/\zZ}(E))=  i2\pi \int_{\wW/\zZ}{\widehat A}
(\Omega(\wW/\zZ)){\rm ch}\,(\Omega(E))_{(2)}\, \in \,\Omega^2(\zZ)\,,
\end{equation}
where $\Omega(\wW/\zZ)$ and $\Omega(E)$ are the curvature forms.

Now it is possible to apply this geometric setup to compute the holonomy on $\zZ$.
Let $\partial\yY\rightarrow {S}^1$ be a loop of manifolds. A metric and spin structure on 
$\partial\yY$ could be induced by
a metric and bounding spin structure on ${S}^1$  . The holonomy of the
determinant line bundles around the loop takes the form
\begin{equation}
{\bf hol}\,{\rm det}\,{\mD}_{\partial\yY/{S}^1}(E)={\bf a}\!-\!{\rm lim}
\tau_{\partial\yY}^{-1}(E)\,,
\label{alim}
\end{equation}
where {\bf a}--{\rm lim} is the adiabatic limit, i.e. the limit as the metric on 
${S}^1$ blows up ($\varepsilon\rightarrow 0:\,\, {g}_{{S}^1}
\rightarrow {g}_{{S}^1}\varepsilon^{-2}$). For the flat determinant 
line bundles no adiabatic limit is required. A nontrivial result for (\ref{alim}) means that
the determinant line bundle (fermion determinant, see below) is nontrivial, which is 
tantamount to saying that there is a global anomaly. Eq. (\ref{alim}) is in fact known 
as the global anomaly formula \cite{witten,bismut1,bismut2}.

Let us summarize. The differential geometry of determinant line bundles 
has been developed in \cite{quillen} in a special case and in \cite{bismut1,bismut2} 
in general. In  \cite{Dai,freed} the results on $\xi$-invariants 
were used to re-demonstrate the holonomy formula for determinant line bundles, known as 
Witten's global anomaly formula \cite{witten}. 
For a family of Dirac operators the exponentiated $\xi$-invariant is
a section of the inverse determinant line bundle over the parameter space. 
In \cite{Dai} the usual formula for the variation of the $\xi$-invariant was been generalized 
to a formula for the covariant derivative. The variational formula relates the exponentiated 
$\xi$-invariant to the natural connection on the (inverse) determinant line bundle \cite{Dai}. 
One can use such a connection to compute the holonomy, or global anomaly. The latter  can expressed as 
the adiabatic limit of the exponentiated $\xi$-invariant. 
 
Returning to Chern-Simons invariants, we have seen that the latter are strictly connected
to the $\eta$ invariants and to the global anomaly formula. In the
original papers, Atiyah, Patodi, and Singer discuss the relationship of $\eta$-invariants 
(and so exponentiated $\xi$-invariants) to classical Chern-Simons invariants for closed manifolds. 
It has been shown \cite{Atiyah77,Atiyah78,Atiyah79} that certain ratios of exponentiated 
$\xi$-invariants are topological invariants which live in $K^{-1}$-theory with 
${\mathbb R}/{\mathbb Z}$ coefficients.
The exponentiated $\xi$-invariant is local and therefore it can serve as an action for a field 
theory, the same one can say for the Chern-Simons invariant. 
But there is also a crucial difference: the Chern-Simons invariant is multiplicative 
in coverings, whereas the exponentiated $\xi$-invariant is not (nevertheless the gluing law 
does exhibit some local properties of the $\eta$-invariant).

The last considerations lead us to the physical interpretation of the material collected so far.

{\bf Note on fermion theories.}
In theories containing fermions interacting with a gauge potential $A$, by
formally integrating out the fermionic fields, one gets an expression
which is interpreted as the determinant of the corresponding Dirac operator (fermion determinant).
One of the most important problems in quantum field theory is the definition of such a determinant.
 In some cases they are ill-defined due to anomalies. In a three-dimensional manifold $\xX$
we can assume that there are no local anomalies and
we have only to worry about global anomalies. Formal calculations give for the determinant, 
as a result of the fermion integration, 
the exponential of a term precisely proportional to the CS action for $A$, see \cite{BCLPS}
\footnote{Where the result is obtained by considering a theory of a massive fermion and taking 
the mass to 0.} and references therein. Thus the fermion determinant is well defined if 
this exponential is and this is so if condition {\rm (\ref{sequential})} is satisfied. For 
instance, this global anomaly vanishes if the number of integrated out Dirac fermions is even 
(or the number of integrated out Majorana fermions are a multiple of four). This, however, is not enough.
Since $\exp \left(2i\pi  CS({\underline A}_{\rho})\right)$ must have the same value whatever 
is the manifold $\MM$ 
over which we perform the integral {\rm (\ref{CS0'})}, Eq. {\rm (\ref{gluing})} requires
\begin{equation}
\exp(i\pi\eta(\nN))=1
\label{witten}
\end{equation}
for any three-manifold $\nN$ without boundary. This condition, relying on the Dai-Freed theorem, 
has been analyzed in {\rm \cite{Witten1}}. Whether a theory satisfies or not {\rm (\ref{witten})} 
depends on the 
number of fermions and the type of fermions in it, i.e. whether they are Majorana or Dirac. 

The same anomaly (the so-called parity anomaly) also originates from the presence of massless
 Majorana fermions on the three-dimensional space $\xX$. They give rise to a determinant 
line bundle which leads precisely to
the calculation outlined above. The result shows up in the form of the $\eta$ invariant 
(or, better, the $\tau$  invariant), but the analysis is parallel to the previous one and leads to the same conclusions.

In \cite{Witten1} the previous results are used to analyze the 3+1 dimensional theories 
that describe
the so-called topological insulators and topological superconductors. These theories are 
defined on a 3+1 manifold
with a boundary and   it is usually necessary to enforce complementarity between the 
fermions in the bulk and those in the boundary in order to cancel the global anomalies 
and satisfy the condition analogous to   {\rm (\ref{witten})}.

\subsection{Adiabatic limit and twisted spectral functions}
\label{Unitary}

Suppose that $\xX = \Gamma\backslash \overline{\xX}$ with $\overline{\xX}$ a globally 
symmetric space of non-compact type and $\Gamma$ a discrete,  torsion-free, co-compact 
subgroup of orientation-preserving isometries. Thus $\xX$ inherits a locally symmetric
Riemannian metric $g$ of non-positive sectional curvature. In addition the connected 
components of the periodic set of the geodesic flow $\Phi$, acting on the unit tangent 
bundle $T\xX$, are parametrized by the non-trivial conjugacy classes $[\gamma]$ in
$\Gamma = \pi_1(\xX)$. Therefore each connected component $\xX_\gamma$ is itself a 
closed locally symmetric manifold of non-positive sectional curvature.

Suppose that $\varphi: \Gamma\rightarrow U({F})$ be a unitary representation of $\Gamma$ on $F$.
The Hermitian vector bundle ${E}= \textsf{X}\times_{\Gamma}{F}$ over $\textsf{X}$
inherits a flat connection from the trivial connection on
$\overline{\xX}\times {F}$. For any vector bundle $E$ over $\xX$ let $\overline{E}$ denote
the pull-back to $\overline{\xX}$.
If ${\mD}: C^{\infty}(\textsf{X}, V)\rightarrow C^{\infty}(\textsf{X},V)$ is a differential
operator acting on the sections of the vector bundle $V$, then
${\mathfrak D}$ extends canonically to a differential operator
${\mathfrak D}_{\varphi}: C^{\infty}(\textsf{X},V\otimes F)\rightarrow
C^{\infty}(\textsf{X},V\otimes{F})$, uniquely
characterized by the property that ${\mathfrak D}_{\varphi}$ is
locally isomorphic to ${\mathfrak D}\otimes \cdots \otimes {\mathfrak
D}$\,\,\, {\rm (}${\rm dim}\,{F}$ times{\rm )}.

\begin{example}
The formula {\rm (\ref{CS1})} suggests the generalization of the Chern-Simons invariant 
{\rm (\ref{CS0})} to the case of non-trivial $U(n)$-bundle over $\xX$. As an example, 
for any representation $\rho: \Gamma \rightarrow U(n)$, a vector bundle $\underline{E}_\rho$ 
over a certain four-manifold $\MM$ with $\partial\MM = \xX = S^3/\Gamma$ {\rm (}which is an 
extension of a flat vector bundle $E_\rho$ over $S^3/\Gamma${\rm )} has been constructed in 
{\rm \cite{Nishi}}.
In the case $\underline{A}_\rho$ is any extension of a flat connection $A_\rho$
corresponding to $\rho$ the index theorem for the twisted Dirac operator ${\mathfrak D}_\rho$
is given by {\rm (}Cf. Eq. {\rm (\ref{Index}))}
\begin{equation}
{\bf Index}\, {\mathfrak D}_{\rho} =\int_\textsf{M}
{\rm ch}
({ {\underline   E}}_{\rho})
{\widehat  A}(\textsf{M})- \sum_{\gamma\neq 1}
\frac{\chi_\rho (\gamma)}{\vert\Gamma\vert(2-\chi_r(\gamma))}\,.
\label{IndexS3}
\end{equation}
Here $r$ denotes the two-dimensional representation induced by the inclusion 
$\Gamma\rightarrow SU(2)$, $\chi$ denotes the character of a representation, while 
$\vert\Gamma\vert$ is the order of $\Gamma$ {\rm (}see {\rm \cite{KN,Nishi}} for 
more detail{\rm )}.
\end{example}

In connection with a real compact hyperbolic manifold $\xX$ consider a locally homogeneous
Dirac bundle $E$ over $\xX$ and the corresponding Dirac operator 
$\mD: C^\infty(\xX, E)\rightarrow C^\infty(\xX, E)$.
As before, assume that $\xX = \partial\MM$, that $E$ extends to a Clifford bundle on
$\MM$ 
\footnote{  
A Clifford module bundle is called a Dirac bundle if it has a connection
$\nabla$ satisfying the compatibility condition $\nabla_z(v\cdot s) = (\nabla_z^Rv)
\cdot s + v\cdot (\nabla_z s)$. Here $s$ is a local section of $E$, $v$ is a local 
section of $C\ell(\MM)$, $z$ a vector field and $(\cdot)$ denote the module multiplication.
On a Dirac bundle one then has a Dirac operator defined by $\mD s = \sum_je_j\cdot
(\nabla_{e_j}s)$, where $\{e_j\}$ is any local orthonormal frame for $\MM$.
} 
and that $\varphi: \pi_1(\xX)\rightarrow U(F)$ extends to a representation of
$\pi_1(\MM)$. Let $\underline{A}_\varphi$ be an extension of a flat connection $A_\varphi$
corresponding to $\varphi$.

{\bf The Cayley transform determinant and adiabatic limit.}
Let us consider a determinant construction for a self-adjoint operator 
on a finite dimensional Hilbert space.
The classical Cayley transform \cite{Cayley}  
for such an operator $\mD$ is the unitary operator $C = (\mD-i)/(\mD + i)$. 
For $s\in {\mathbb C}$ we have a family of operators
\begin{equation}
C(s) = \frac{\mD-is}{\mD+is}\,. 
\end{equation}
This family is meromorphic, has poles at $s\in i{\rm Spec}^{\prime}({\mathfrak D})$\,
$({\rm Spec}^{\prime}(\mD)\equiv{\rm Spec}({\mathfrak D})-\{0\})$, these poles being simple 
and having residue ${\rm Res}_{-i\lambda}C(s) = 2i\lambda P_\lambda$, where $P_\lambda$
is projection onto the $i\lambda$ eigenspace. One has (see for detail \cite{Moscovici89}):
\begin{equation}
{\rm log}\,{\rm det}^\prime C(s) = \!\sum_{\lambda \in {\rm Spec}^\prime (\mD)}
m(\lambda){\rm log} \left(\frac{\lambda -is}{\lambda+ is}\right), 
\end{equation}
where $m(\lambda)$ denote the multiplicity. 

Let $\mD$ be a Dirac operator, as defined above; the 
family of operators $C(s)= (\mD)-is)/(\mD+is)$ is meromorphic with simple poles at
$s\in i{\rm Spec}^\prime (\mD)$. The determinant satisfies the functional identity 
$
{\rm det}^\prime(\mD+ is)/(\mD-is)\cdot {\rm det}^\prime (\mD-is)/(\mD+ is) = 1.
$
The following result holds \cite{Moscovici89} (Proposition 2.2):
\begin{equation}
\lim_{\stackrel{x\rightarrow +\infty}{}}
\, {\rm det}^\prime\, C(x) = e^{-i\pi\eta(0, \mD)}.
\label{adiabatic1}
\end{equation}
Set $\varepsilon = x^{-1}$. Then if one replaces the metric $g$ on $\xX$ by 
$g_\varepsilon = g\varepsilon^{-1}$ then 
Eq. {\rm (\ref{adiabatic1})} says that the adiabatic limit ({\bf a}--{\rm lim}) of the Cayley 
transform of $\mD_\varepsilon$ is $\exp (-i\pi\eta(0, \mD))$ (Cf. Eq. (\ref{alim}))
\begin{equation}
{\bf a}\!-\!\lim_{\stackrel{\varepsilon\rightarrow 0}{}}
\, {\rm det}^\prime\, \frac{\mD - i\varepsilon^{-1}}{\mD +i\varepsilon^{-1}} 
= e^{-i\pi\eta(0, \mD)}.
\label{adiabatic2}
\end{equation}

{\bf Locally symmetric spaces of higher rank.}
It has been shown \cite{F1,F2} that for variety flows the zeta function associated to any cyclic 
flat bundle is actually meromorphic on a neighborhood of $[0, \infty)$, regular at $s=0$, 
and its value at $s=0$ coincides with R-torsion with coefficients in the given flat bundle, 
and thus is a topological invariant. Recall that Ray and Singer defined an analytic torsion 
$\tau_\rho^{\bf an}(\xX)\in (0, \infty)$ for every closed Riemannian manifold $\xX$ and 
orthogonal representation $\rho: \pi_1(\xX)\rightarrow O(n)$ \cite{RS}.
Because of the analogy with the Lefschetz fixed point formula, Fried proved that the geodesic 
flow of a closed manifold of constant negative curvature has the Lefschetz property 
\cite{F3}. He also conjectured that this remains true for any closed locally homogeneous 
Riemannian manifold.

Fried's conjecture has been proved and an adequate theory of Selberg-type zeta functions 
for locally symmetric spaces of higher rank was constructed in \cite{Moscovici91}. 
Difficulties have been avoided by constructing certain {\it super} Selberg zeta functions,
$Z^\ell(s, \mD_\varphi)$, $0\leq \ell\leq 2m< {\rm dim}\,\xX$, as alternating products
of formal Selberg-like functions, which reduce to Selberg zeta functions only in the
three-dimensional rank one case. Each function $Z^\ell(s, \mD_\varphi)$ is 
meromorphic on $\mathbb C$ and moreover satisfies a functional equation (see \cite{Moscovici91} 
for details). Not surprisingly, the functional equations  play a crucial role in identifying 
the special value of the Selberg-type spectral function ${\cR}(s; \varphi)$ with the R-torsion. 
Finally, 
${\cR}(s; \varphi)$ can be expressed as an alternating product of $Z^\ell$,
\begin{equation}
{\cR}(s; \varphi) = \prod_{\ell =0}^{{\rm dim}\,\xX -1} Z^\ell(s-{\rm dim}\,\xX+1+\ell ,
\,\mD_\varphi)^{(-1)^\ell}\,.
\end{equation}
On this basis we conjecture that the Chern-Simons invariant for locally symmetric spaces 
of higher rank admits 
a representation in terms of the spectral function ${\cR}(s; \varphi)$. 
This is an interesting and important conclusion and we hope to come back to this analysis 
in the future.

\section{Infinite products for the quantum $\mathfrak{sl}_N$ invariant}
\label{QuantumCS}

In this section we consider more general correlators in a CS theory, 
and view them as generating series of quantum group invariants weighted by S-functions. 
The quantum group invariants can be defined over any semi-simple Lie algebra $\mathfrak g$.
In the $SU(N)$ Chern-Simons gauge theory we study the quantum ${\mathfrak s}{\mathfrak l}_N$ 
invariants, which can be identified as the so-called colored HOMFLY polynomials
\footnote{
The framed HOMFLY polynomial of links (an invariant of framed oriented links), is denoted
by $\cH(\cL)$, and can be normalized as follows:
$\cH(\bigcirc) = (t^{-\frac12}-t^{\frac12})/(q^{-\frac12}-q^{\frac12})$.
(These invariants can be recursively computed through the HOMFLY skein.)
The colored HOMFLY polynomials are defined through the \emph{satellite knot}. A satellite of knot 
$\cK$ is determined by choosing a diagram $Q$ in the annulus. Draw $Q$ on the annular 
neighborhood of $\cK$ determined by the framing, to give a satellite knot $\cK\star Q$.
One can refer to this construction as \emph{decorating $\cK$ with the pattern $Q$}. The HOMFLY
polynomial $\cH(\cK\star Q)$ of the satellite depends on $Q$ only as an element of the skein
$\cC$ of the annulus. $\{Q_\lambda\}_{\lambda\in\cY}$ form a basis of $\cC$. $\cC$ can be
regarded as the parameter space for these invariants of $\cK$, and can be  called
the HOMFLY {\it satellite invariants of $\cK$.}
}

One important corollary of the LMOV conjecture is the possibility to express 
a Chern-Simons partition function as an infinite product. In this article we derive such a product. 
During the calculations we use the characters of the symplectic groups. The latter were
found by Weyl \cite{Weyl} using a transcendental method
(based on integration over the group manifold). However the appropriate characters may 
also be obtained by algebraic methods \cite{Littlewood44}. Following \cite{Fauser10} we have 
used algebraic methods. This allows to exploit the Hopf algebra methods to determine
(sub)group branching rules and the decomposition of tensor products.

The motivation for studying an infinite-product formula, associated to topological string 
partition functions, based on a guess on the modular property of partition function, 
stimulated by properties of S-functions. 

{\bf Preliminaries.}
To derive the infinite-product formula, we need some preliminary material. 
First of all we denote by $\cY$ the set of all Young diagrams. Let $\chi_A$ 
be the character of the irreducible
representation of the symmetric group labeled by a partition $A$. Given a partition $\mu$,
define $m_j = \textrm{card} (\mu_k=j; k\geq 1)$. (The order of the conjugate class of type
$\mu$ is given by: $\mathfrak{z}_\mu = \prod_{j\geq1} j^{m_j} m_j!.$)
The symmetric power functions of a given set of variables $X=\{x_j\}_{j\geq 1}$
are defined as the direct limit of the Newton polynomials:
$p_n(X) = \sum_{j\geq1} x_j^n, \, \, p_\mu(X) = \prod_{i\geq 1} p_{\mu_i}(X),$
and we have the following formula which determine the Schur function and the orthogonality
property of the character
\begin{equation}
s_A(X) = \sum_{\mu} \frac{ \chi_A(C_\mu) }{\mathfrak{z}_\mu} p_\mu(X), \,\,\,\,\,\,\,\,\,
\sum_{\mu} \frac{ \chi_A(C_\mu) \chi_B(C_\mu) }{ \mathfrak{z}_\mu } = \delta_{A,B}\,.
\end{equation}
where $C_\mu$ denotes the conjugate class of the symmetric 
{group $S_{\vert \mu \vert}$} 
corresponding to partition $\mu$ (for details see Sect. 3 of [33]).

Given $X= \{x_i\}_{i\geq 1}$, $Y=\{y_j\}_{j\geq 1}$, define
$
X\ast Y = \{x_i\cdot y_j\}_{i\geq 1, j\geq 1}.
$
We also define $X^d = \{ x_i^d\}_{i\geq 1}$.
The $d$-th Adams operation of a Schur function is given by $s_A(X^d)$. 
(An Adams operation is type of algebraic construction; the basic idea of this operation 
is to implement some fundamental identities in S-functions. In particular, $s_A(X^d)$ 
means operation of a power sum on a polynomial.)

We use the following conventions for the notation:
\begin{itemize}
\item{} 
$\cL$ will denote a link and $L$ the number of components in $\cL$.
\item{}
The irreducible $U_q(\mathfrak{sl}_N)$ module associated to $\mathcal L$ will be labeled by
their highest   weights, thus by Young diagrams. We usually denote it by a vector form
$\overrightarrow{A}=(A^1,\ldots,A^L)$.
\item{}
Let $\overrightarrow{X} =(x_1,\ldots,x_L)$ be a set of $L$  variables,
each of which is associated to  a component of $\mathcal L$ and
$\overrightarrow{\mu} = (\mu^1,\ldots,\mu^L)\in\cY^L$ be a tuple of $L$ partitions. We write:
$$
[\overrightarrow{\mu}] = \prod_{\alpha=1}^L [\mu^\alpha],
\,\,\,\,\,\,\, \mathfrak{z}_{\overrightarrow{\mu}} = \prod_{\alpha=1}^L \mathfrak{z}_{\mu^\alpha},
\,\,\,\,\,\,\, \chi_{\overrightarrow{A}}(C_{\overrightarrow{\mu}}) =
\prod_{\alpha=1}^L \chi_{A^\alpha}(C_{\mu^\alpha}),
$$
$$
s_{\overrightarrow{A}}(\overrightarrow{X}) =
\prod_{\alpha=1}^L s_{A^\alpha}(x_\alpha), \,\,\,\,\,\,\,
p_{\mu }(X)=\overset{\ell (\mu )}{\prod_{i=1}}p_{\mu _{i}}(X),\,\,\,\,\,\,\,
p_{\overrightarrow{\mu}}(\overrightarrow{X}) = \prod_{\alpha=1}^L p_{\mu^\alpha}(x_\alpha).
\nonumber
$$
\end{itemize}

{\bf The case of links and a knot.}
The quantum $\mathfrak{sl}_N$ invariant for the irreducible module $V_{A^1},\ldots,V_{A^L}$,
labeled by the corresponding partitions $A^1,\ldots, A^L$,  can be
identified as the HOMFLY invariants for the link decorated by $Q_{A^1},\ldots,Q_{A^L}$.
The quantum $\mathfrak{sl}_N$ invariants of the link is given by
$
P_{\overrightarrow{A}}(\mathcal{L}; q,t) =
\mathcal{H} (\mathcal{L}\star \otimes_{\alpha=1}^L Q_{A^\alpha} ).
$
The colored HOMFLY polynomial of the link $\mathcal L$ can be defined by \cite{Zhu}
\begin{equation}
P_{\overrightarrow{A}} = q^{-\sum_{\alpha = 1}^Lk_{A^\alpha}\omega({\mathcal K}_\alpha)}
t^{-\sum_{\alpha = 1}^L \vert A^\alpha\vert \omega({\mathcal K}_\alpha)}
\langle \mathcal{L}\star \otimes_{\alpha=1}^L Q_{A^\alpha} \rangle\,,
\end{equation}
where $\omega({\mathcal K}_\alpha)$ is the number of the $\alpha$-component
${\mathcal K}_\alpha$ of $\mathcal L$
and the bracket $\langle \mathcal{L}\star \otimes_{\alpha=1}^L Q_{A^\alpha} \rangle$ denotes
the framed HOMFLY polynomial of the satellite link
$\mathcal{L}\star \otimes_{\alpha=1}^L Q_{A^\alpha}$.
We can define the following invariants:
\begin{equation}\label{definition of W_mu}
\textsf{W}_{\overrightarrow{\mu}}(\mathcal{L};q,t) =
\sum_{\overrightarrow{A} =(A^1,\ldots,A^L) } \bigg( \prod_{\alpha=1}^L
\chi_{A^\alpha}(C_{\mu^\alpha} ) \bigg)
P_{\overrightarrow{A}}(\mathcal{L};q,t)\,.
\end{equation}

The Chern-Simons partition function $\textsf{W}^{SL}_{CS}({\mathcal L};q,t)$ and the free energy
$F({\mathcal L};q,t)$ of the link ${\mathcal L}$ are the following generating series of
quantum group invariants weighted by Schur functions $s_{\overrightarrow{A}}$ and by the
invariants $\textsf{W}_{\overrightarrow{\mu}}$:
\begin{eqnarray}
\textsf{W}^{SL}_{CS}({\mathcal L};q,t) & = &
1+ \sum_{\overrightarrow{A}} P_{\overrightarrow{A}} ({\mathcal L}; q,t)
s_{\overrightarrow{A}}(\overrightarrow{X}) =
1+ \sum_{\overrightarrow{\mu} }
\frac{ \textsf{W}_{\overrightarrow{\mu}}
({\mathcal L};q,t) }{{\mathfrak z}_{\overrightarrow{\mu}} }
p_{\overrightarrow{\mu}}(\overrightarrow{X}) \,,
\label{CS-A}
\\
F({\mathcal L};q,t) & = & \log \textsf{W}_{CS}({\mathcal L};q,t)
= \sum_{\overrightarrow{\mu}}
\frac{ F_{\overrightarrow{\mu}}({\mathcal L};q,t) }{{\mathfrak z}_{\overrightarrow{\mu}}}
p_{\overrightarrow{\mu}}(\overrightarrow{X}) \,.
\end{eqnarray}

Based on LMOV conjecture the infinite product formula for the case of links, 
$\textsf{W}_{CS}^{SL}({\mathcal L};q,t; \overrightarrow{X})$ and a knot 
$\textsf{W}_{CS}^{SL}(\cK;q,t;X)$ are given by \cite{Liu2,LiuPeng}  
\begin{eqnarray}
\textsf{W}_{CS}^{SL}({\mathcal L};q,t; \overrightarrow{X}) & = &
\prod_{\overrightarrow{\mu}}\,
\prod_{Q\in {\mathbb Z}/2} \, \prod_{m=1}^\infty\prod_{k= -\infty}^\infty\,
\big\langle 1-  q^{k+m}t^Q \, \overrightarrow{X}\big\rangle^{-n_{\overrightarrow{\mu};\,g,Q}},
\label{SP1}
\\
\textsf{W}_{CS}^{SL}(\cK;q,t;X) & = &   
\prod_{\mu}\prod_{Q\in {\mathbb Z}/2} \,\, \prod_{m=1}^\infty \;
\prod_{k = -\infty}^\infty\;
\big \langle 1- q^{k+m}t^Q   X^\mu
\big \rangle^{-m\, {n}_{\mu;\,g,Q}}\,.
\label{SP2}
\end{eqnarray} 
Here $\overrightarrow{\mu}=(\mu^1,\ldots,\mu^L)$, the length of $\mu^i$ is $\ell_i$, 
$\overrightarrow{X}=(x_1,\ldots, x_L)$, and 
${n}_{\mu;\,g,Q}$ are invariants related to the integer invariants in the LMOV
conjecture. For a given $\mu$, ${n}_{\mu;g,Q}$ vanish for sufficiently 
large $|Q|$ due to the vanishing property of $n_{\mu;\,g,Q}$. The products involving $Q$ 
and $k$ are finite products for a fixed partition $\mu$.

The symmetric product $\big \langle 1- q^{k+m}t^Q   X^\mu \big \rangle$ and the generalized 
symmetric product $\big\langle 1-  q^{k+m}t^Q \, \overrightarrow{X}\big\rangle$  in Eqs. 
(\ref{SP2}) and (\ref{SP1}), respectively, are defined by the formula \cite{LiuPeng}
\begin{eqnarray}
\big\langle 1- \psi\, X^\mu \big\rangle & = & 
\prod_{ x_{i_1},\ldots, x_{i_{\ell(\mu)} } }
\Big( 1- \psi\, x_{i_1}^{\mu_1} \cdots x_{i_{\ell(\mu)}}^{\mu_{\ell(\mu)}} \Big)
\label{Angle1},
\\
\big\langle 1-\psi\, \overrightarrow{X}^{{\mu}}\big\rangle & = &   
\prod_{\alpha=1}^L\prod_{i_{\alpha, 1},\ldots, i_{\alpha, \ell_\alpha}}
\Big(1- \psi\prod_{\alpha=1}^L
\big((x_\alpha)_{ i_{\alpha,1} }^{\mu^\alpha_1} \cdots
(x_\alpha)_{ i_{\alpha, \ell_\alpha} }^{\mu^\alpha_{\ell_\alpha}}\big)\Big)\,.
\label{Angle2}
\end{eqnarray}
where $\psi$ is a generic variable.

{\bf Double series and certain modular forms.}
{In Eqs. (\ref{SP1}), (\ref{SP2}) 
and (\ref{Angle1}), (\ref{Angle2}) the blocks for affine-like denominators admit representations 
in the form of spectral functions of hyperbolic geometry (see for example \cite{BBC}). 
One can successfully construct quantum homological invariants and express the formal character 
of the irreducible tensor representations of the classical groups in terms of the symmetric 
and spectral functions of hyperbolic geometry \cite{BC}. 
Indeed, he products in Eqs. (\ref{SP1}) and (\ref{SP2}) can be represented in the form of
Selberg-type spectral function $\cR(s)$ of hyperbolic three-geometry.} ${\mathcal R}(s)$ 
is an alternating product of more complicated factors, each of which is a so-called 
Patterson-Selberg
zeta-functions $Z_{\Gamma^\gamma}$ \cite{Bytsenko1,Bytsenko2} ($\cR(s)$ can be continued 
meromorphically to the entire complex plane $\mathbb C$),
\begin{eqnarray}
\prod_{n=\ell}^{\infty}(1- q^{an+\varepsilon})
& = & \prod_{p=0, 1}Z_{\Gamma^\gamma}(\underbrace{(a\ell+\varepsilon)(1-i\varrho(\vartheta))
+ 1 -a}_s + a(1 + i\varrho(\vartheta)p)^{(-1)^p}
\nonumber \\
& = &
\cR(s = (a\ell + \varepsilon)(1-i\varrho(\vartheta)) + 1-a),
\label{R1}
\\
\prod_{n=\ell}^{\infty}(1+ q^{an+\varepsilon})
& = &
\prod_{p=0, 1}Z_{\Gamma^\gamma}(\underbrace{(a\ell+\varepsilon)(1-i\varrho(\vartheta)) + 1-a +
i\sigma(\vartheta)}_s
+ a(1+ i\varrho(\vartheta)p)^{(-1)^p}
\nonumber \\
& = &
\cR(s = (a\ell + \varepsilon)(1-i\varrho(\vartheta)) + 1-a + i\sigma(\vartheta))\,,
\label{R2}
\end{eqnarray}
where $q= \exp(2\pi i\vartheta)$, $\varrho(\vartheta) =
{\rm Re}\,\vartheta/{\rm Im}\,\vartheta$,
$\sigma(\vartheta) = (2\,{\rm In}\,\vartheta)^{-1}$,
$a$ is a real number, $\varepsilon, b\in {\mathbb C}$, $\ell \in {\mathbb Z}_+$.
In terms of $\cR(s)$ functions, Eq. (\ref{R1}), (see also 
\cite{BC}, Eq (3.41)) 
$\textsf{W}_{CS}^{SL}({\mathcal L};q,t; \overrightarrow{X})$ and 
$\textsf{W}_{CS}^{SL}(\cK;q,t;X)$ take the form
\begin{eqnarray}
\!\!\!\!\!\!\!\!\!\!\!\!\!\!\!\!\!\!\!\!\!\!\!\!\!\!
&&
\textsf{W}_{CS}^{SL}({\mathcal L};q,t; \overrightarrow{X}) =
\nonumber \\
\!\!\!\!\!\!\!\!\!\!\!\!\!\!\!\!\!\!\!\!\!\!\!\!\!\!
&&
\prod_{\overrightarrow{\mu}}\,
\prod_{Q\in {\mathbb Z}/2} \,\prod_{k= -\infty}^\infty\,\prod_{\alpha=1}^L\
\prod_{i_{\alpha, 1},\ldots, i_{\alpha, \ell_\alpha} }\prod_{m=1}^\infty
{\mathcal R}\left(s= (m+\Omega(q^{g-2k}t^Q\overrightarrow{X}^\mu; \vartheta))
(1-i\varrho(\vartheta))\right)^{-n_{\overrightarrow{\mu};\,g,Q}}\!\!\!,
\label{Link}
\\
\!\!\!\!\!\!\!\!\!\!\!\!\!\!\!\!\!\!\!\!\!\!\!\!\!\!
&&
\textsf{W}_{CS}^{SL}(\cK;q,t;X) =
\nonumber \\
\!\!\!\!\!\!\!\!\!\!\!\!\!\!\!\!\!\!\!\!\!\!\!\!\!\!
&&
\prod_{\mu}\prod_{Q\in {\mathbb Z}/2} \,
\prod_{k = -\infty}^\infty \,\prod_{ x_{i_1},\ldots, x_{i_{\ell(\mu)} }}\, \prod_{m=1}^\infty
\left({\mathcal R}(s= (m+\Omega(t^QX^\mu q^k;\vartheta))
(1-i\varrho(\vartheta)))\right)^{-{n}_{\mu;\,g,Q}},
\label{Knot}
\end{eqnarray}
where, $\Omega(q^kt^QX^\mu; \vartheta)\equiv 
{\rm log}(q^kt^Qx_{i_1}^{\mu_1}\cdots x_{i_{\ell(\mu)}}^{\mu_{\ell(\mu)}})/2\pi i\vartheta$.

Calculations in the case of Kauffman polynomials, relative to the orthogonal group, can be 
found in a recent paper \cite{BC}. 

To finish we consider one more topic of interest, the symmetries and the modular form 
identities. 
Both Hecke \cite{Hecke} and Rogers \cite{Rogers} 
recognized that certain modular forms could be represented by combinations of the 
following double series:
\begin{equation}
\sum_{(m,n)\in  {\Omega}}(-1)^{{\mathcal H}(m,n)}q^{{\mathcal L}(m,n) + {\mathcal Q}(m, n)}\,.
\label{double}
\end{equation}
Here  ${\mathcal H},\, {\mathcal L}$ are linear forms, ${\mathcal Q}$ is an indefinite 
quadratic form and $\Omega$ is some subset of ${\mathbb Z}\times {\mathbb Z}$. 
\footnote{We also mention on this topic the deep results of
\cite{Kac80,Kac-Peterson}, where a number of identities in the representation 
theory of Ka\v{c}-Moody Lie algebras has been obtained. A family of modular functions 
satisfying Rogers-Ramanujan type identities for arbitrary affine root systems has been 
obtained in \cite{Cherednik}. Extensive work in the theory of partition identities 
shows that basic hypergeometric series provide the generating functions for numerous 
families of partition identities.}

Infinite double series of type (\ref{double}) and their connection with the $\cR(s)$ 
function have been investigated in \cite{BBC}.  For the following functions 
(see \cite{BBC} for detail)
\begin{eqnarray}
Q_{n, k}(a_1, a_2, a_3, a_4, a_5; q) & := & (-1)^{n+k}q^{a_1n+a_2k+a_3nk+a_4n^2+a_5k^2}\,,
\label{Q}
\\
{\mathcal J}(a_1, a_2, a_3, a_4, a_5; q) & := &
\left(\sum_{n, k\geq 0} - \sum_{n, k< 0}\right)
Q_{n, k}(a_1, a_2, a_3, a_4, a_5; q)
\label{SQ}
\end{eqnarray}
one can find an infinite family of identities. 
\begin{eqnarray}
\sum_{n,k = -\infty;\, k\geq |2n|}^\infty (-1)^{n+k}q^{(k^2-3n^2)/2+(n+k)/2}
& = & \prod_{n= 1}^\infty (1-q^n)^2
\label{Rogers}
\\
\sum_{n,k = -\infty;\, k\geq |2n|}^\infty (-1)^{n+k}q^{(k^2-3n^2)/2+(n+k)/2}
& \stackrel{(\ref{SQ}),\, (\ref{R1})}{=\!=\!=\!=\!=\!=}&
\!\!\! \sum_{n, k = -\infty}^\infty Q_{n, k\geq\vert 2n\vert}
\left(\frac{1}{2}, \frac{1}{2}, 0, -\frac{3}{2}, \frac{1}{2}; q\right)
\nonumber \\
& = &
\cR(s=1-i\varrho(\vartheta))^2\,.
\label{DE0}
\end{eqnarray}
The identity (\ref{Rogers}) was conjectured by Rogers  and has been proved in 
\cite{Hecke,Kac-Peterson}.
We finally remark that Rogers' approach can be used to discover possible modular 
identities and symmetry properties of the infinite products considered in this paper
(the simplest symmetry is $q\rightarrow q^{-1}$), by using connections between 
Hecke-Rogers modular form identities and functional equations for $\cR(s)$.

\section*{Acknowledgments}

A. A. B. would like to acknowledge the Conselho Nacional de Desenvolvimento 
Cient\'{i}fico e Tecnol\'{o}gico (CNPq, Brazil) and Coordena\c{c}\~{a}o de Aperfei\c{c}amento 
de Pessoal de N\'{i}vel Superior (CAPES, Brazil) for financial support. L.B. would like to thank
the Departamento de Fisica da Universitade Estadual de Londrina for its kind hospitality.

\end{document}